# *PERELMAN*: Pipeline for scientific literature meta-analysis. Technical report

**Daniil Sherki[1,2] , Daniil Merkulov[1,2] , Alexandra Savina[2] , Ekaterina Muravleva[1,2]**

[1]AI4Science Center, Sber, [2]Skoltech

We present **PERELMAN** (PipEline foR sciEntific Literature Meta-ANalysis), an agentic framework designed to extract specific information from a large corpus of scientific articles to support large-scale literature reviews and meta-analyses. Our central goal is to reliably transform heterogeneous article content into a unified, machine-readable representation. **PERELMAN** first elicits domain knowledge—including target variables, inclusion criteria, units, and normalization rules—through a structured dialogue with a subject-matter expert. This domain knowledge is then reused across multiple stages of the pipeline and guides coordinated agents in extracting evidence from narrative text, tables, and figures, enabling consistent aggregation across studies. In order to assess reproducibility and validate our implementation, we evaluate the system on the task of reproducing the meta-analysis of layered Li-ion cathode properties $LiNi_{0.8}Mn_{0.1}Co_{0.1}O_2$ (NMC811) reported in [1]. We describe our solution, which has the potential to reduce the time required to prepare meta-analyses from months to minutes.

**Date:** December 25, 2025
**Version:** 0.1

# Contents

# 1 Introduction

At present, the volume of scientific literature is growing at an unprecedented rate. Preprint platforms such as arXiv alone release thousands of new submissions annually, with the number of published papers in 2025 being approximately twice that of 2019 [2]. Beyond preprints, the Crossref DOI registry—covering journal and conference publishing—reports metadata for 178 million records, including >120 million journal-article DOIs and 9.6 million conference-proceedings DOIs (updated Dec 25, 2025) [3]. This rapid expansion makes it increasingly difficult for researchers to systematically track new results and maintain an up-to-date understanding of their fields.

At the same time, many research areas require continuous monitoring of domain-specific knowledge and experimental practices. For instance, in lithium-ion battery research, hundreds of studies are published each year on the properties of individual cathode materials, making manual synthesis and comparison of results increasingly impractical.

As a result, the construction of comprehensive and reproducible meta-analyses has become a critical challenge for the scientific community across a wide range of domains, including chemistry, medicine, materials science, physics, and machine learning.

We present a pipeline for agentic processing of large amounts of data from scientific papers. We call our pipeline **PERELMAN** (PipEline foR sciEntific Literature Meta-ANalysis). This is an agentic pipeline whose main goal is the extraction of target data from a large corpus of domain-specific articles for subsequent expert-driven meta-review writing. **PERELMAN** is capable of:
(i) extracting domain-specific knowledge through interaction with experts (in development);
(ii) searching for and collecting scientific literature within a specific domain (in development);
(iii) extracting data using a Vision Language Model (VLM) agent capable of using tools for image normalization (cropping and rotation) and identifying the target image;
(iv) finding parameters in the article text using an LLM.

For demonstration purposes, we used our agentic pipeline to reproduce data from a meta-review in the electrochemistry domain [1]. We also work on other experimental data domains, but they were not included in the current technical report. We further probe a central challenge for multi-modal extraction: a large fraction of relevant values appear only in graphical materials, where current VLMs are known to underperform human experts, especially in complex scientific plots [4].

Our contributions can be summarized as follows:

- We propose **PERELMAN**, an agentic framework for large-scale scientific literature reviews that integrates article collection, robust multi-modal parsing, and VLM agent-based data extraction.
- We implemented an agentic system based on article parsing using `docling` ([5], [6]) and `granite` [7], with a Qwen-based VLM agent [8].
- Using the NMC811 benchmark of [1] as a reference, we quantitatively validate the framework against a strong VLM-only baseline, and evaluate each part of the pipeline.

## 2 Pipeline

### 2.1 Agentic System Architecture

The architecture of the agentic system consists of three major components:
(i) agentic system for extracting domain knowledge (in development),
(ii) an article-parsing tool,
(iii) an agent for extracting data from the parsed articles, using domain knowledge.

Principal agentic system architecture is presented at Figure 1.

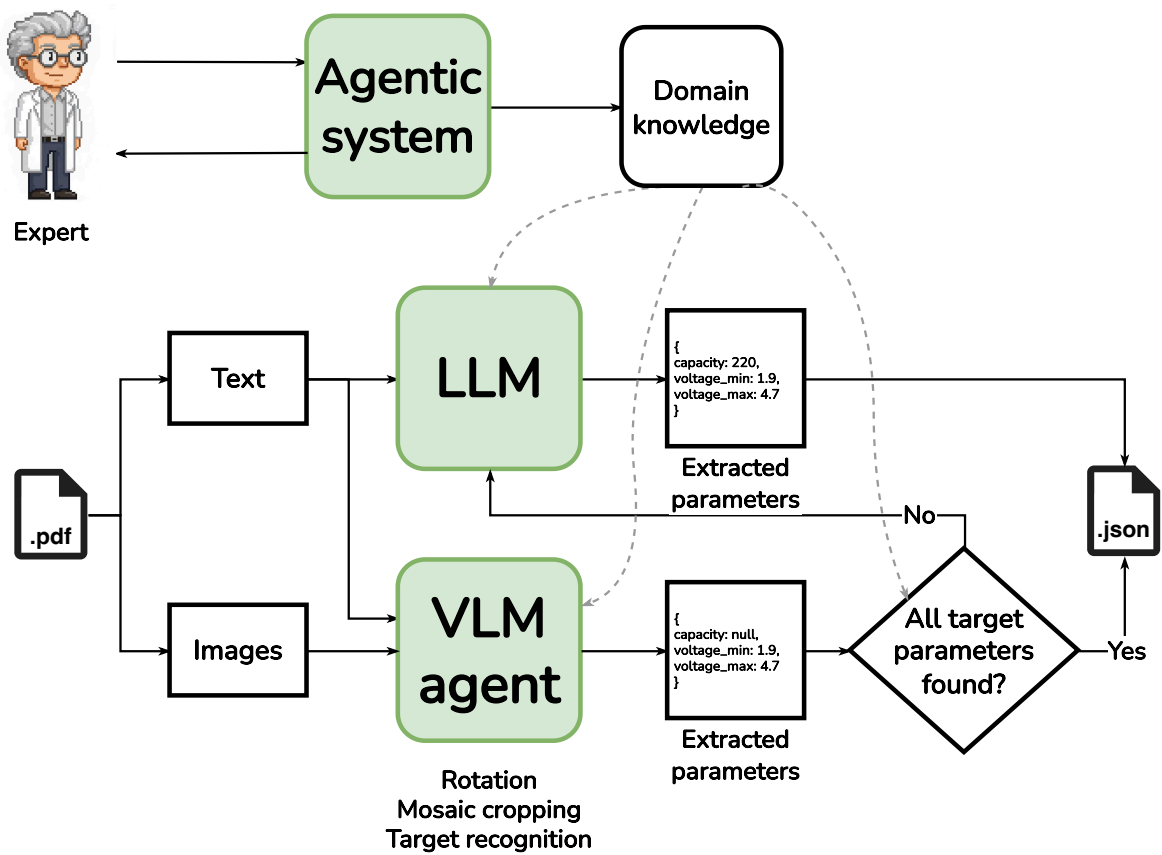

Figure 1: Principal PERELMAN Architecture scheme

### 2.2 Domain knowledge

The goal of this component of the agentic system is to acquire domain knowledge through an interactive dialogue with a human expert and subsequently apply it to the analysis of the collected scientific literature. Domain knowledge is leveraged throughout all stages of intelligent information processing: to frame the problem, identify target variables, design agent prompts, and guide downstream reasoning. In Figure 1, the integration of domain knowledge into the system is illustrated by dashed arrows.

### 2.3 Data Collection

The article-collection tool is currently under development and was not included in the first release of PERELMAN. For the purpose of this technical report, and specifically for the experi-ments in Section 3, a relevant subset of 24 articles was manually curated from the dataset of [1] in collaboration with a domain expert. The target figures from which data are extracted were annotated by hand, and the failure modes and bottlenecks encountered during the execution of the pipeline were systematically analyzed.

## 2.4 Article Parsing

The core tool for article parsing may be any OCR system capable of processing tables and converting them into an LLM-readable format. In our case, the best results were obtained with `Docling` [5], [6]. We convert each PDF into Markdown with layout-aware OCR and extract figure images into a dedicated folder, preserving figure captions when available. The Markdown is stored with referenced images (not embedded) to keep the text stream compact while retaining links to the original figures.

The processing of graphical materials from the articles is handled as a separate stage. In our framework, the VLM model [7] is used exclusively for Optical Character Recognition (OCR), converting the visual content of the articles into an LLM-readable Markdown representation.

## 2.5 VLM Agent

### 2.5.1 Figure data extraction

Data extraction is multimodal. For each figure we build an extra context: the caption plus a +/- 1000 character window around figure mentions in the Markdown. This context is passed to the vision model to ground extraction in nearby textual descriptions.

### 2.5.2 Figure Targeting and Mosaic Handling Normalization

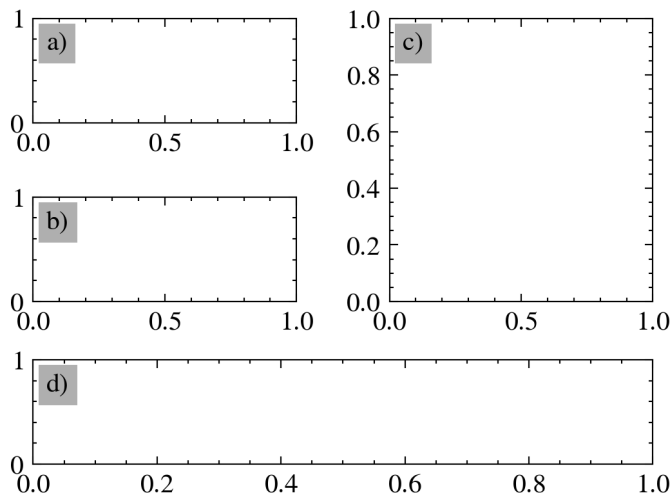

The VLM model represents this image as a structured `mosaic` encoding, which is internally perceived as follows.

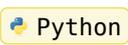

```Python
1    '''
2    ac
3    bc
4    dd
5    '''
```

Figure 2: The example of subplot mosaic representation

We use a VLM to predict a strict JSON schema that includes the mosaic grid (rows/cols) as Figure 2 example shown, panel labels, an optional target panel, whether any panel is relevant, and required rotation. If the model outputs a 1x1 grid but labels imply multiple panels, we adjust the grid; if no labels are present, we apply a simple aspect-ratio heuristic to split tall or wide mosaics. Then the images are uniformly split into panels.

Within the same model invocation, we additionally predict one of the rotation angles required to orient the image (0°, 90°,180°,270°). The figure is rotated if needed using the tools available to the agent.

### 2.5.3 Multi-modal Metric Extraction

After image normalization, we infer a large VLM model, which takes the target image as input, receives the system prompt Section A.1.2, and the full **extra content** described in Section 2.5.1. The target parameters are then extracted in `json` format, as specified during the expert dialogue and stored in **domain knowledge** (Section 2.2). If the agentic system fails to extract data from the target images, we fall back to a Large Language Model (LLM); in our default setup, the text-only extractor uses `GPT-oss 120B` [9]. We run text-only extraction over the Markdown using the prompt in Section A.1.1. The Markdown is chunked with overlap, and the first non-null values are merged into the final record.

### 2.5.4 Outputs

At the end of the pipeline, we obtain a `json` file containing the extracted target metrics, which is stored in the output database for subsequent expert analysis and meta-review writing.

## 3 Experiments

### 3.1 NMC811 reference benchmark

Layered oxides of the form $LiNi_xMn_yCo_zO_2$ (NMC, $x + y + z = 1$) are among the most important cathode materials for Li-ion batteries. Within this family, the Ni-rich composition $LiNi_{0.8}Mn_{0.1}Co_{0.1}O_2$ (NMC811) offers high specific capacity ($\approx 200$ mAh g$^{-1}$) and is a leading candidate for high-energy automotive cells [10]. At the same time, Ni-rich NMC materials suffer from complex degradation mechanisms associated with oxygen release, surface reconstruction, and electrolyte reactivity [11]. As a result, the literature on NMC811 spans hundreds of studies with partially inconsistent electrochemical metrics and experimental conditions.

A. Savina and A. Abakumov recently addressed this by performing a large-scale manual meta-analysis of NMC811 as a positive electrode material, aggregating 548 articles and more than 950 experimental records into benchmark distributions for First cycle Discharge Capacity, First cycle Coulombic efficiency, and cycling stability [1]. Their work demonstrates both the value of such meta-analyses and the substantial human effort required to produce them.

These distributions provide benchmark values for NMC811 and related compositions and are widely used as reference points for new cathode optimization studies. We replicate as closely as possible the document set used in the NMC811 benchmark study. For each record, the reference meta-analysis reports:

- first-cycle discharge capacity (mAh g$^{-1}$),
- first-cycle Coulombic efficiency (%),
- upper cut-off voltage and C-rate,
- additional descriptors (e.g., synthesis route, particle morphology).

# 4 Results

## 4.1 NMC811 meta-analysis validation

For the validation of our framework, we used the articles collected in [1]. As a baseline, we employed a VLM (Qwen, [12]) and performed page-by-page inference on the article by splitting it into chunks and applying the same prompt used by our agent to extract the required parameters.

The comparison results with the baseline are presented in Table 1. We also investigated the impact of different configuration settings. For example, the effect of the VLM size used for chart annotation on the overall performance of the framework is shown in Table 2. The metric reported is the accuracy of data extraction across articles (in percent). It can be observed that using a model with a larger number of parameters improves the overall accuracy of the pipeline.

| System | First Discharge Capacity | Coulombic Efficiency | Voltage (max) | Voltage (min) |
|---|---|---|---|---|
| **Ours** | **58.33%** | **33.33%** | **95.83%** | 70.80% |
| GPT-5.2 Pro Extended Thinking | 25.00% | 0.00% | 79.20% | **79.20%** |
| GPT-5.2 Thinking Heavy | 8.30% | 8.30% | 58.30% | 58.30% |

Table 1: Accuracy by type for different systems

In this specific setup, the cropping module achieves an accuracy of 70.83%, meaning that in this fraction of cases the target image is correctly cropped without being corrupted. The target-image classifier reaches an accuracy of 88.25% given that the image has been correctly cropped. Overall, the effective performance of the Image Normalizer module is therefore $70.83\% \cdot 88.25\% = 62.5\%$.

| VLM | Text | First cycle Discharge Capacity | Coulombic efficiency | Voltage (max) | Voltage (min) |
|---|---|---|---|---|---|
| — | gpt-oss-120b | 19.23% | 11.54% | 42.31% | 30.77% |
| granite-2b | gpt-oss-120b | 50.00% | 19.23% | 80.77% | 69.23% |
| granite-8b | gpt-oss-120b | 53.85% | 30.77% | 84.62% | 73.08% |

Table 2: Extraction quality by type (%)

According to Table 3, the majority of information sources are contained within the graphical materials of the articles. Furthermore, The performance on chart analysis is significantly lower than on text or tables (in old version of pipeline, parameters extraction quality on text 86%, tables 100%. But on chart only 34.6% Thus, the ability to answer questions about charts is a critically important task.

| Source | Fraction |
|---|---|
| Text | 29% |
| Table | 1% |
| Chart | 70% |

Table 3: Target parameters source type distribution

It is essential not only to accurately annotate charts but also to provide high-quality answers to chart-related questions.

## Chart-based Question Answering

Specialized benchmarks exist for chart question answering. The well-known study [4] demonstrates that the performance of modern VLMs on charts from contemporary scientific papers does not reach human-level accuracy. We selected the top-performing models from this benchmark, as well as several state-of-the-art VLMs, and evaluated their ability to retrieve information from target charts using the prompt Section A.1.1. The results are presented in Table 4.

| Model | Overall | Capacity | Coulombic Efficiency | Voltage (max) | Voltage (min) |
|---|---|---|---|---|---|
| Anthropic Claude 3.7 Sonnet (Vision) [13] | 30.68% | 0.00% | 0.00% | 72.73% | 50.00% |
| Anthropic Claude 4.5 [14] | 0.00% | 0.00% | 0.00% | 0.00% | 0.00% |
| Google Gemma 3 27B IT [15] | 0.00% | 0.00% | 0.00% | 0.00% | 0.00% |
| Zhipu AI GLM 4.5V [16] | 7.95% | 0.00% | 0.00% | 18.18% | 13.64% |
| OpenAI GPT-5 Mini [17] | 9.09% | 0.00% | 0.00% | 13.64% | 22.73% |
| xAI Grok-4 [18] | 3.57% | 0.00% | 0.00% | 7.14% | 7.14% |
| NVIDIA Nemotron-Nano-12B-V2-VL [19] | 15.28% | 0.00% | 0.00% | 27.78% | 33.33% |
| Alibaba Qwen 3 VL 235B A22B Instruct [12] | 17.05% | 4.55% | 0.00% | 31.82% | 31.82% |

Table 4: VLM evaluation results

After prompt-tuning, the prompt was finalized as Section A.1.2. This enables metric extraction exclusively from target images selected by the expert, cropped and normalized. Importantly, prompt-tuning primarily consisted of injecting expert-provided domain knowledge into the prompt: instead of merely instructing the model to locate specific target parameters, we provide precise definitions, explain where these parameters are most commonly reported, in which types of plots they typically appear, and how a human expert would navigate such plots to identify the desired values. As shown in Table 5, enriching the prompt with expert domain knowledge leads to an order-of-magnitude improvement in target-parameter extraction accuracy from images (from 4.55% to 80.95% for one of the parameters).

| Approach | Capacity | Coulombic efficiency | Voltage (max) | Voltage (min) |
|---|---|---|---|---|
| Default prompt w/o domain knowledge | 4.55% | 0.00% | 31.82% | 31.82% |
| After prompt-tuning and image normalization | 80.95% | 36.36% | 95.65% | 73.91% |

Table 5: Accuracy by task type for qwen3-vl-235B (3.0% tolerance)

# 5 Discussion and Limitations

While developing the pipeline, we encountered several key challenges. Even very large VLMs struggle to extract target parameters from images when provided with minimal domain knowledge, as evidenced by the results in Table 2. Incorporating expert-derived domain knowledge has a substantial impact on the overall performance of the pipeline.

A second challenge is that the end-to-end quality of the pipeline is tightly coupled to the performance of each individual component. If the target-image classifier performs poorly, the final pipeline metrics degrade accordingly.

In addition, many figures in scientific articles are complex and composed of multiple subfigures. The current mosaic-based processing often fails in such cases, as it assumes that images can be evenly partitioned, an assumption that does not always hold.

Planned next steps include:
- improving mosaic splitting;
- integrating a module for extracting domain knowledge from expert chat interactions, as well as a module for retrieving relevant literature;
- removing watermarks from figures in scientific articles.

# 6 Related work

Automating evidence synthesis has been most visible in biomedicine, where individual steps of systematic reviews are partially automated. RobotReviewer predicts risk-of-bias assessments and retrieves supporting rationale spans from full texts, while ASReview applies active learning to reduce the number of titles/abstracts that require manual screening [20], [21].

At the infrastructure layer, large machine-readable corpora and PDF parsers make large-scale literature processing feasible. S2ORC provides structured scientific full text with linked mentions of figures and tables, enabling retrieval and benchmarking at scale. Classical pipelines such as GROBID convert scholarly PDFs into structured TEI representations, and PubLayNet provides large-scale page layout annotations for training layout detectors on scientific articles [22], [23], [24].

For scientific natural language processing (NLP), domain-adapted pretraining and supervised IE benchmarks focus primarily on text. SciBERT improves representation learning for scientific language, and SciERC/SciIE formalize joint extraction of entities, relations, and coreference for scientific knowledge graphs. More domain-specific expert-labeled datasets, such as POLYIE for polymer literature, extend these paradigms towards materials-relevant entities, values, and n-ary relations [25], [26], [27].

Multimodal extraction is substantially harder when key results appear only in charts. ChartQA benchmarks visual + logical reasoning over charts, but accurate numerical recovery remains challenging in dense, heterogeneous plots. Recent domain-focused work in materials science curates in-domain chart-to-table benchmarks and introduces evaluation metrics that emphasize coordinate/label fidelity, highlighting the gap between generic chart benchmarks and real scientific figures [28], [29].

Domain-specific chemistry/materials extraction has historically relied on toolkits and rules for recognizing chemical entities and property relations. ChemDataExtractor is a widely used toolkit in this direction and has been applied to build large literature-mined battery-material databases. In parallel, recent evaluations of LLMs for materials-science information extraction report that relation reasoning can be strong with few-shot examples, while robust recognition and normalization of domain-specific entities remains a bottleneck—motivating hybrid pipelines with explicit schemas and provenance [30], [31], [32].

Finally, agentic tool-use methods provide a general template for decomposing complex workflows into interpretable action sequences with intermediate artifacts. ReAct interleaves reasoning with tool calls, and Toolformer learns when to invoke external tools, supporting modular designs where retrieval, parsing, and extraction are separable and auditable [33], [34].

# 7 Conclusions

We introduced **PERELMAN** (PipEline foR sciEntific Literature Meta-ANalysis), an agentic framework for large-scale, multi-modal literature reviews. **PERELMAN** combines modular document collection, layout-aware parsing into an LLM-readable representation, and figure-centric multi-modal extraction with fallbacks to text-only extraction. Using the NMC811 benchmark meta-analysis as a reference task, we demonstrated that an explicit agentic control flow—especially figure targeting, mosaic handling, and chart-aware prompting—substantially improves extraction accuracy over a VLM-only baseline.

Our analysis highlights chart question answering as the dominant bottleneck: most target values are contained in figures, and chart extraction accuracy remains far below text extraction accuracy. Nevertheless, the pipeline already reduces the manual burden by producing auditable sidecar artifacts with provenance, enabling rapid iteration, ablation studies, and selective human verification.

Finally, we demonstrated that the same architecture transfers beyond a single materials system (NMC622) and beyond materials science (LLM training reports) by changing only the extraction schema and prompts. We expect such modular agentic pipelines to become a practical bridge between fast-growing scientific literature and reproducible, continuously updated meta-analyses.

# APPENDIX A Experiment details

## A.1 Prompts

### A.1.1 Extraction from markdown

```
Mosaic Prompt                                                              [⚙ JSON]
 1  You are a precise extractor for scientific battery papers.
 2
 3  Task
 4  Return a STRICT JSON object with these keys (all required, null if not found):
 5  {
 6    "voltage_range": "min-max" | null,
 7    "voltage_min": number | null,
 8    "voltage_max": number | null,
 9    "capacity": number | null,
10    "coulombic_efficiency_pct": number | null
11  }
12
13  Rules
14  - Use ONLY explicit numbers found in the text or tables. Do not guess.
15  - If both min and max are found, also provide "voltage_range" as "min-max".
16  - Decimal separator is dot. No units in values.
17  - If multiple conditions exist, prefer the main experimental condition.
18  - Return ONLY JSON, no extra text.
```

### A.1.2 Prompt for VLM

```
Mosaic Prompt                                                              [⚙ JSON]
 1  You are a vision-language model that extracts numeric data from scientific battery plots for the NMC811/NCM811
    cathode.
 2
 3  About the material & synonyms
 4
 5  - NMC811 is a Ni-rich layered oxide cathode with composition LiNi0.8Mn0.1Co0.1O2 (order of Mn/Co may vary).
 6  - Common synonyms/variants: "NMC", "NCM","MNC","CNM","CMN","MCN", "NMC811", "NCM811", "NMC-811", "NCM-811",
    "811 NMC", "LiNi0.8Mn0.1Co0.1O2", "LiNi0.8Co0.1Mn0.1O2", "Li[Ni0.8Mn0.1Co0.1]O2", "Ni-rich layered oxide
    (NMC811)", "SC-NMC811", "pristine/bare NMC811".
 7  - If the figure compares modifications (coated/doped/with …), the baseline series is the "cleanest" one:
    "NMC811/NCM811" without modifiers like "@…", "coated", "doped", "with …", "gradient", "core-shell".
 8
 9  What to extract
10
11  Return EXACTLY the target metrics below in a strict JSON format (see "Output format").
12  You must rely ONLY on information visible inside the image: main title, panel titles (a/b/c), axis labels,
    legend, in-figure callouts/annotations, captions printed inside the figure bounds. Also, see the description
    text and all figure mentions from appended
13
14  Target metrics to return
15
16  1) Voltage window (Voltage range, V) — as "min-max".
17     - IMPORTANT: For V-Q / Voltage-Capacity (Type A) plots, DO NOT take min/max from axis tick limits.
18     - Instead, derive the voltage window from the FIRST-CYCLE curve endpoints (cut-off voltages) on the target
       curve:
19       * V_max = voltage at the first-cycle end-of-charge point (at C_ch,1).
20       * V_min = voltage at the first-cycle end-of-discharge point (at C_dis,1).
21
22
23  2) First discharge capacity (mAh/g) (first-cycle discharge capacity, C_dis,1).
24
25  3) First charge capacity (mAh/g) (first-cycle charge capacity, C_ch,1).
```

26      - This MUST be extracted from the V–Q / Voltage–Capacity first-cycle curve if Type A is present.

27

28  Output format (strict order, strict schema)

29

30  Return ONLY the following JSON array (no extra text, no Markdown):

31

32  [
33    {
34      "type": "voltage",
35      "question": "Voltage range (V)",
36      "answer": "<min>-<max>" | null,
37      "correct_answer": null
38    },
39    {
40      "type": "capacity",
41      "question": "First discharge capacity (mAh/g)",
42      "answer": number | null,
43      "correct_answer": null
44    },
45    {
46      "type": "charge_capacity",
47      "question": "First charge capacity (mAh/g)",
48      "answer": number | null,
49      "correct_answer": null
50    }
51  ]

52

53  Hard rules

54

55  - Return ONLY one JSON array exactly as above (no extra text, no Markdown).
56  - "answer" is:
57    * a string "min-max" for voltage, OR
58    * a NUMBER for capacity / charge_capacity / FCE.
59  - If a value is not reliably determinable, return null for that value.
60  - Keep "correct_answer" present but null.
61  - Decimal separator is dot; round ALL numeric answers to ONE decimal place.
62  - Do not infer anything not visible on the image.
63  - CRITICAL ACCURACY RULES:
64    * Extract EXACT values from the plot/data. Do NOT round to "nice" values unless that is exactly what is shown.
65    * Read precise numbers from axes, data points, or text labels.
66    * NEVER default to round numbers (e.g., 200.0, 208.0) unless the plot explicitly shows them.

67

68  Determine the chart type

69

70  - Type A (V–Q / Voltage–Capacity / Voltage–Specific Capacity):
71    * One axis is Voltage (V) / Potential (V) (often y-axis).
72    * The other axis is capacity: "Specific capacity (mAh/g)", "Capacity (mAh g^-1)", "Q (mAh g^-1)", etc.
73    * Often shows multiple cycles (e.g., "1st", "2nd", "10th") and/or charge vs discharge branches.
74  - Type B (non Voltage–Capacity):
75    * Examples: capacity vs cycle number; rate performance; CE vs cycle; bar chart/table of capacities; etc.

76

77  Rules for Type A (V–Q / Voltage–Capacity)

78

79  STEP-BY-STEP INSTRUCTIONS:

80

81  0) Choose the correct series (baseline NMC811)
82     - If multiple materials exist, choose baseline/pristine/bare NMC811 (no coating/doping/modifier).
83     - If baseline cannot be identified, return null for all metrics that depend on series selection.

84

85  1) Identify the FIRST cycle curve
86     - Use legend labels like "1st cycle", "Cycle 1", "1st", "First".

```
87        - If unclear, pick the curve explicitly representing the earliest cycle.
88        - If still ambiguous, return null for first-cycle-dependent metrics.
89
90   2) Identify charge vs discharge branches (first cycle)
91        - Charge branch: voltage increases as the process proceeds from 0 capacity to max (left to right).
92        - Discharge branch: voltage decreases as the process proceeds from 0 capacity to max (left to right).
93        - Do NOT assume "rightmost" always means end; some plots may be drawn left-to-right or right-to-left.
94        - Use the voltage trend (increasing vs decreasing) and endpoint voltages to identify ends.
95
96   3) Extract first-cycle charge capacity (C_ch,1) — MUST OUTPUT
97        - Find the END-OF-CHARGE point on the first-cycle CHARGE branch:
98          * This is where the first-cycle curve reaches its HIGHEST voltage (upper cut-off, V_max) and terminates
             for charge.
99        - Read the capacity value (mAh/g) at that endpoint from the capacity axis.
100       - This is C_ch,1 → output in the "charge_capacity" object.
101
102  4) Extract first-cycle discharge capacity (C_dis,1)
103       - Find the END-OF-DISCHARGE point on the first-cycle DISCHARGE branch:
104         * This is where the first-cycle curve reaches its LOWEST voltage (lower cut-off, V_min) and terminates for
            discharge.
105       - Read the capacity value (mAh/g) at that endpoint from the capacity axis.
106       - This is C_dis,1 → output in the "capacity" object.
107
108  5) Extract voltage window (V) — CRITICAL FIX (DO NOT USE AXIS TICK LIMITS)
109       - DO NOT use the minimum/maximum tick marks of the voltage axis.
110       - Instead, read voltages from the FIRST-CYCLE CURVE ENDPOINTS:
111         * V_max = voltage at end-of-charge point (the same point used for C_ch,1).
112         * V_min = voltage at end-of-discharge point (the same point used for C_dis,1).
113       - Read these two voltages from the voltage axis scale at those endpoints.
114       - Output as "V_min-V_max" (rounded to one decimal place, e.g., "2.8-4.3").
115
116  Rules for Type B (non Voltage—Capacity)
117
118  1) Voltage range (V)
119       - Search ONLY in textual elements inside the figure bounds.
120       - Look for patterns like "2.8-4.3 V", "2.5-4.5 V", "Cut-off: 2.7-4.4 V".
121       - Extract exact numbers shown; if not present, return null.
122       - Do NOT confuse with current densities or C-rates.
123
124  2) Baseline series selection
125       - Identify baseline/pristine/bare NMC811 as described above.
126       - If ambiguous, return null for metrics that depend on series selection.
127
128  3) First discharge capacity (mAh/g)
129       - If there is a capacity-vs-cycle plot:
130         * Find the DISCHARGE series and read the value at cycle 1.
131       - If only one capacity series is shown and not labeled:
132         * Assume it is discharge capacity and read cycle 1 (earliest point).
133       - Extract the exact value from axis/labels; if not readable, return null.
134
135  4) First charge capacity (mAh/g)
136       - Only if a CHARGE series is explicitly present (or table values show charge at cycle 1):
137         * Read charge capacity at cycle 1 for the baseline.
138       - Otherwise return null for charge_capacity.
139
140  Additional rules
141
142  - Units: capacity must be in mAh/g.
143       * If axis shows "Ah/g", multiply by 1000 (e.g., 0.198 Ah/g = 198.0 mAh/g).
144       * If units are unclear or missing, return null for capacity metrics.
145  - If the figure does not include NMC811/NCM811 at all, return null for all metrics.
146
```

```
147  Critical mistakes to avoid (must not happen)

148

149  - DO NOT report voltage window using the voltage-axis tick limits in Type A plots.

150    * Voltage window must come from the first-cycle curve endpoints (end-of-charge and end-of-discharge
         voltages).

151  - DO NOT confuse charge and discharge branches.

152  - DO NOT use values from the wrong cycle or wrong material variant.

153

154  Final requirement

155

156  Return ONLY the single JSON array of three objects in the exact format above. Be precise, extract values from
       the figure, and compute FCE only from the extracted first-cycle charge/discharge capacities.
```

### A.1.3 Prompt for mosaic extraction

**Mosaic Prompt**  [⚙ JSON]

```
1   You are a precise vision model. Given ONE scientific figure (often multi-panel), return ONLY a JSON object with
    these keys:

2   {

3     "rows": int,              // grid rows (>=1)

4     "cols": int,              // grid cols (>=1)

5     "panel_labels": [string], // optional, e.g., ["a","b","c"]

6     "target_panel": string|null,// label or sequential id of the panel that contains voltage—capacity or capacity-
                                   vs-cycle (first-cycle info)

7     "is_target": 0|1,         // 0 if ANY panel has extractable first-cycle capacity/voltage; 1 otherwise

8     "rotation": 0|90|180|270  // clockwise rotation needed to read text upright

9   }

10  Rules:

11  - rows/cols: describe the main mosaic grid that fits all subplots (e.g., 1x2, 2x2). If single plot, rows=1,
      cols=1.

12  - panel_labels: if letters are printed (a,b,c,...) list them in reading order (left-to-right, top-to-bottom).
      Otherwise omit or [].

13  - target_panel: pick the single panel where first-cycle capacity/voltage is most likely readable. Use the
      printed letter if available; otherwise a 1-based index in reading order ("1", "2", ...). If none, null.

14  - is_target: 0 means AT LEAST one panel is relevant; 1 means none are relevant.

15  - rotation: clockwise angle so that text reads left-to-right.

16  Return ONLY JSON, no markdown, no text.
```